\documentclass[11pt]{article}

\usepackage[preprint]{acl}

\usepackage{times}
\usepackage{latexsym}
\usepackage{booktabs}
\usepackage{longtable}
\usepackage{array}
\usepackage{comment}

\usepackage{tikz}
\usetikzlibrary{arrows.meta, positioning, fit, backgrounds}

\definecolor{raggray}{HTML}{F3F4F6}
\definecolor{inputblue}{HTML}{E8F1FF}
\definecolor{diagnosisblue}{HTML}{DCEBFF}
\definecolor{repairpurple}{HTML}{F0E8FF}
\definecolor{humanamber}{HTML}{FFF4D6}
\definecolor{verifygreen}{HTML}{E8F7EE}
\definecolor{darkline}{HTML}{3A3A3A}

\usepackage{tikz}
\usetikzlibrary{shapes.geometric}

\definecolor{inputblue}{HTML}{E8F1FF}
\definecolor{diagnosisblue}{HTML}{DCEBFF}
\definecolor{repairpurple}{HTML}{F0E8FF}
\definecolor{humanamber}{HTML}{FFF4D6}
\definecolor{verifygreen}{HTML}{E8F7EE}
\definecolor{darkline}{HTML}{3A3A3A}

\usepackage{booktabs}
\usepackage{tabularx} 
\usepackage{url}

\usepackage{array} 
\newcolumntype{Y}{>{\raggedright\arraybackslash}X}

\usepackage[T1]{fontenc}

\usepackage[utf8]{inputenc}

\usepackage{microtype}

\usepackage{inconsolata}
\usepackage{fontawesome5}
\usepackage{hyperref}
\usepackage{graphicx}
\usepackage{hyperref}
\newcommand{\squishlist}{
   \begin{list}{$\bullet$}{
      \setlength{\itemsep}{0pt}
      \setlength{\parsep}{3pt}
      \setlength{\topsep}{3pt}
      \setlength{\partopsep}{0pt}
      \setlength{\leftmargin}{1.5em}
      \setlength{\labelwidth}{1em}
      \setlength{\labelsep}{0.5em}
   }
}
\newcommand{\squishend}{\end{list}}

\title{\textcolor{black}{\textsc{Rectify}: An Interactive Workbench for Post-Evaluation RAG Diagnosis, Repair, and Verification}}

\author{
  \textbf{Keerthana Murugaraj},
  \textbf{Salima Lamsiyah},
  \textbf{Martin Theobald} \\
 University of Luxembourg \\ Esch-sur-Alzette, Luxembourg
   \small{
 \textbf{Correspondence:} \href{mailto:keerthana.murugaraj@uni.lu}{keerthana.murugaraj@uni.lu}
 }
}

\begin{document}
\maketitle

\begin{abstract}

\textcolor{black}{
Retrieval-Augmented Generation (RAG) evaluators can identify failures such as weak retrieval, poor grounding, incomplete answers, and unsupported generation, but they rarely help developers decide what to repair next. We present \textsc{Rectify}, an interactive Streamlit workbench that turns evaluated RAG cases into auditable repair workflows. \textsc{Rectify} filters cases that do not require repair, routes remaining failures into actionable families and fine-grained repair slices, and generates editable repair cards that developers can approve, reject, or verify through sandbox reruns. On a controlled RAG benchmark, \textsc{Rectify} surfaces interpretable failure profiles across BM25, dense, and hybrid retrieval: BM25 mainly triggers noisy-retrieval repairs, while dense and hybrid retrieval leave smaller sets of multi-part under-retrieval and underused-evidence cases. Additional analyses show that pre-filtering reduces unnecessary repair candidates and that slice-level routing yields more targeted repair cards than broad family-level diagnosis. \textsc{Rectify} is publicly available as an open-source Streamlit workbench \footnote{\faGithub\ \href{https://github.com/KeerthanaMurugaraj/Rectify-An-Interactive-Workbench-for-Post-Evaluation-RAG-Repair-and-Verification}{Code Repository}} for helping developers turn evaluation results into inspectable repair decisions.
}
\end{abstract}

\section{Introduction}
RAG has become a common design pattern for building language-model systems that answer questions over external documents \citep{lewis2020rag}. This makes RAG useful in settings such as enterprise question answering, scientific search, legal and historical archives, and customer-support systems. \citep{wiratunga2024cbr,Murugaraj_Lamsiyah_During_Theobald_2025, xu2024retrieval} However, RAG pipelines remain difficult to debug: a low-quality answer may arise from failures in retrieval, grounding, answer synthesis, or abstention \citep{es2023ragas,murugaraj2026ragvue,ru2024ragchecker}. Recent surveys highlight that RAG evaluation must account for both retrieval and generation behavior, including relevance, faithfulness, answer quality, robustness, and benchmark design aligned with real user needs \citep{gao2023retrieval,yu2024evaluation}. Yet evaluation alone does not close the debugging loop. Although recent diagnostic frameworks provide actionable per-case recommendations \citep{cohen2025ragxplain}, developers are still often left to decide which failures share a common root cause, which pipeline component should be changed, and whether a proposed change actually improves the affected examples. This post-evaluation step is especially important because treating every failed case as an isolated error makes repair slow, inconsistent, and difficult to audit.

\begin{figure*}[t]
\centering
\resizebox{1.03\linewidth}{!}{
\begin{tikzpicture}[
    node distance=0.55cm,
    every node/.style={font=\footnotesize},
    stage/.style={
        draw=darkline,
        rounded corners=4pt,
        align=center,
        minimum width=2.85cm,
        minimum height=1.12cm,
        line width=0.5pt
    },
    inputstage/.style={stage, fill=inputblue},
    diagnosisstage/.style={stage, fill=diagnosisblue},
    repairstage/.style={stage, fill=repairpurple},
    humanstage/.style={stage, fill=humanamber},
    verifystage/.style={stage, fill=verifygreen},
    logstage/.style={
        draw=darkline,
        rounded corners=3pt,
        align=center,
        minimum width=2.65cm,
        minimum height=0.72cm,
        line width=0.45pt,
        fill=raggray
    },
    arrow/.style={-{Latex[length=2mm]}, line width=0.7pt, draw=darkline},
    dashedarrow/.style={-{Latex[length=2mm]}, line width=0.7pt, draw=darkline, dashed}
]

\node[inputstage] (schema) {
    \textbf{1. Unified Case}\\
    \textbf{Representation}\\
    {\scriptsize answers, contexts, scores}
};

\node[diagnosisstage, right=of schema] (diagnosis) {
    \textbf{2. Failure Families}\\
    \textbf{\& Repair Slices}\\
    {\scriptsize pre-filters; 4 families; 23 slices}
};

\node[repairstage, right=of diagnosis] (cards) {
    \textbf{3. Repair-Card}\\
    \textbf{Generation}\\
    {\scriptsize template-based repair hypotheses}
};

\node[humanstage, right=of cards] (approval) {
    \textbf{4. Human Approval}\\
    \textbf{\& Provenance}\\
    {\scriptsize inspect, edit, approve/reject}
};

\node[verifystage, right=of approval] (sandbox) {
    \textbf{5. Optional Sandbox}\\
    \textbf{Verification}\\
    {\scriptsize rerun cases; report deltas}
};

\node[logstage, below=0.45cm of approval] (log) {
    {\scriptsize provenance log}\\
    {\scriptsize scope, notes, timestamp}
};

\draw[arrow] (schema) -- (diagnosis);
\draw[arrow] (diagnosis) -- (cards);
\draw[arrow] (cards) -- (approval);
\draw[dashedarrow] (approval) -- (sandbox);
\draw[arrow] (approval) -- (log);

\end{tikzpicture}
}
\caption{\textsc{Rectify} post-evaluation workflow. Evaluated cases are normalized, routed into failure families and repair slices, converted into editable repair cards, reviewed by a developer, and optionally verified through sandbox reruns with provenance logging.}
\label{fig:rectify-workflow}
\end{figure*}
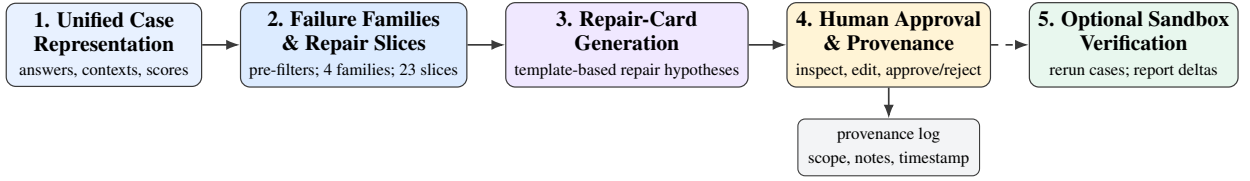

We present \textsc{Rectify}, an interactive workbench for RAG failure diagnosis, repair, and optional verification. \textsc{Rectify} uses RAGVue \citep{murugaraj2026ragvue} as its primary evaluator and turns diagnostic evaluation outputs into a post-evaluation repair workflow. It takes evaluated cases, groups recurring failures into four actionable families: \textit{retrieval}, \textit{grounding}, \textit{generation}, and \textit{abstention}, and maps them to fine-grained repair slices. Each slice is converted into an editable repair card that a developer can inspect, approve, or reject before any change is tested. Approved repairs can optionally be verified in a sandbox on the affected cases, with before and after deltas and provenance logs supporting auditability. Our key contributions are as follows:

\textcolor{black}{\squishlist
    \item We present \textsc{Rectify}, an open-source Streamlit workbench for post-evaluation RAG debugging that turns evaluator outputs into case exploration, failure diagnosis, repair-card review, provenance logging, and sandbox verification.
    \item We introduce a deterministic failure-routing scheme that filters non-actionable cases and maps remaining failures to four macro families and 23 fine-grained repair slices, enabling cluster-level repair decisions rather than case-by-case inspection.
    \item We design editable repair cards that connect each failure slice to inspectable configuration-level interventions over retrieval, chunking, reranking, prompting, abstention, and generation while keeping developers in control of approval and scope.
    \item We evaluate \textsc{Rectify} on a controlled RAG benchmark across BM25, dense, and hybrid retrieval, showing interpretable retriever-specific repair agendas, fewer unnecessary repair candidates after pre-filtering, and more targeted cards than family-level routing.
\squishend}

\section{Related Work}

\paragraph{RAG Evaluation and Benchmarks.}
Recent work has developed metrics and benchmarks for evaluating RAG systems beyond end-task accuracy. RAGAS introduced reference-free metrics for faithfulness, answer relevancy, context precision, and context recall \citep{es2023ragas}. ARES trains lightweight judges for context relevance, answer faithfulness, and answer relevance using synthetic data and limited human annotation \citep{saadfalcon2023ares}. RAGChecker separates retrieval and generation behavior through fine-grained diagnostic metrics \citep{ru2024ragchecker}. RAGVue provides diagnostic and explainable reference-free evaluation across retrieval quality, answer relevance and completeness, strict faithfulness, and calibration \citep{murugaraj2026ragvue}. Complementary benchmarks study hallucination, robustness, citation-supported generation, and actionable evaluation labels, including RAGTruth, RGB, ALCE, and RAGBench \citep{niu2024ragtruth,chen2023rgb,gao2023alce,friel2024ragbench}. These works expose important quality signals, but they do not by themselves define a human-approved repair workflow.

\paragraph{From Evaluation to Guidance and Repair.}
Recent approaches move from evaluation toward developer guidance. RAGXplain converts evaluation scores into per-case natural-language explanations for improving RAG pipelines \citep{cohen2025ragxplain}. RAGGY provides composable RAG primitives and an interactive interface for real-time pipeline debugging \citep{lauro2025raggy}. ARAGOG compares RAG configurations such as reranking, multi-query retrieval, maximal marginal relevance, HyDE, and sentence-window retrieval \citep{eibich2024aragog}. Doctor-RAG studies failure-aware repair for agentic RAG by localizing failures in reasoning trajectories and repairing the diagnosed point \citep{jiao2026doctorrag}. These systems make evaluation more actionable, but they do not focus on dataset-level post-evaluation failure grouping, human approval of repair cards, and measured before--after verification for standard RAG pipelines.


In contrast to the existing works, \textsc{Rectify} addresses the gap between RAG evaluation and pipeline revision. It is neither another evaluator nor an automatic self-repair system: it starts from already evaluated cases and treats proposed changes as repair hypotheses that require developer inspection and approval. Given evaluator outputs, \textsc{Rectify} groups recurring failures into repair slices, generates editable configuration-level repair cards, optionally verifies approved repairs on the affected cases, reports before and after deltas, and records decisions in a provenance log. This positions \textsc{Rectify} as a post-evaluation workbench for auditable, human-controlled RAG diagnosis and repair.

\section{The \textsc{Rectify} System}

\textsc{Rectify} operates after a RAG pipeline has produced answers and an evaluator has scored them. Figure~\ref{fig:rectify-workflow} summarizes the post-evaluation workflow.

\subsection{Unified Case Representation}

\textsc{Rectify} first normalizes each evaluated RAG case into a shared schema. Each case contains the question, generated answer, retrieved contexts, optional expected answer, evaluator scores, diagnostic fields, and available metadata. This common representation allows \textsc{Rectify} to apply the same workflow across cases: failure routing, repair-card generation, human review, provenance logging, and optional sandbox verification. In the current implementation, RAGVue is the primary evaluator because it provides diagnostic signals for retrieval quality, grounding, answer completeness, abstention behavior, and response quality. These signals are used to assign cases to failure families and fine-grained repair slices.


\subsection{Failure Families \& Repair Slices}\label{main:failure_family}

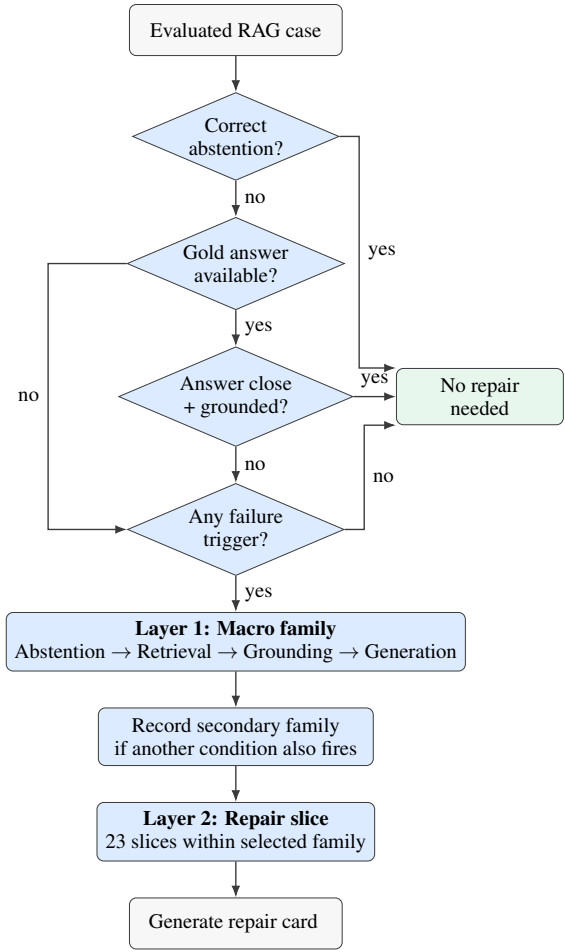
\begin{figure}[t]
\centering
\resizebox{0.98\linewidth}{!}{
\begin{tikzpicture}[
    node distance=0.55cm,
    every node/.style={font=\footnotesize},
    box/.style={
        draw=darkline,
        rounded corners=3pt,
        align=center,
        minimum width=3.25cm,
        minimum height=0.78cm,
        fill=gray!6
    },
    healthybox/.style={
        draw=darkline,
        rounded corners=3pt,
        align=center,
        minimum width=2.55cm,
        minimum height=0.86cm,
        fill=verifygreen
    },
    decision/.style={
        draw=darkline,
        diamond,
        aspect=2.35,
        align=center,
        inner sep=1.6pt,
        minimum width=2.70cm,
        minimum height=0.95cm,
        fill=diagnosisblue
    },
    layer/.style={
        draw=darkline,
        rounded corners=3pt,
        align=center,
        minimum width=4.20cm,
        minimum height=0.88cm,
        fill=diagnosisblue
    },
    arrow/.style={-{Latex[length=1.8mm]}, line width=0.6pt, draw=darkline}
]
\node[box] (case) {Evaluated RAG case};
\node[decision, below=of case] (abstain) {
    Correct\\abstention?
};
\node[decision, below=of abstain] (gold) {
    Gold answer\\available?
};
\node[decision, below=of gold] (goldfilter) {
    Answer close\\+ grounded?
};
\node[decision, below=of goldfilter] (trigger) {
    Any failure\\trigger?
};
\node[healthybox, right=0.65cm of goldfilter] (skip) {
    No repair\\needed
};
\node[layer, below=of trigger] (layer1) {
    \textbf{Layer 1: Macro family}\\
    Abstention $\rightarrow$ Retrieval $\rightarrow$ Grounding $\rightarrow$ Generation
};
\node[layer, below=of layer1] (secondary) {
    Record secondary family\\
    if another condition also fires
};
\node[layer, below=of secondary] (layer2) {
    \textbf{Layer 2: Repair slice}\\
    23 slices within selected family
};
\node[box, below=of layer2] (card) {
    Generate repair card
};
\draw[arrow] (case) -- (abstain);
\draw[arrow] (abstain) -- node[right] {no} (gold);
\draw[arrow] (gold) -- node[right] {yes} (goldfilter);
\draw[arrow] (goldfilter) -- node[right] {no} (trigger);
\draw[arrow] (trigger) -- node[right] {yes} (layer1);
\draw[arrow] (abstain.east) -- ++(0.30,0)
    |- node[pos=0.25,right] {yes} (skip.north west);
\draw[arrow] (goldfilter.east) -- node[above] {yes} (skip.west);
\draw[arrow] (trigger.east) -- ++(0.30,0)
    |- node[pos=0.25,right] {no} (skip.south west);
\draw[arrow] (gold.west) -- ++(-1.20,0)
    |- node[left,pos=0.25] {no} (trigger.west);
\draw[arrow] (layer1) -- (secondary);
\draw[arrow] (secondary) -- (layer2);
\draw[arrow] (layer2) -- (card);
\end{tikzpicture}
}
\caption{Failure routing in \textsc{Rectify}.}
\label{fig:failure-routing}
\end{figure}

\textsc{Rectify} uses a two-layer taxonomy to convert evaluator outputs into repairable failure patterns. Before routing begins, two pre-filters remove cases that do not require repair. First, a correct-abstention filter excludes unanswerable cases when the model appropriately refuses to answer. Second, a gold-answer filter excludes answerable cases when the generated answer is sufficiently close to the gold answer and grounded in the retrieved context. The routing process is illustrated in Figure~\ref{fig:failure-routing}.

\begin{table*}[t]
\centering
\scriptsize
\setlength{\tabcolsep}{4pt}
\renewcommand{\arraystretch}{1.05}
\begin{tabularx}{\textwidth}{p{0.13\textwidth}p{0.06\textwidth}p{0.23\textwidth}X}
\toprule
\textbf{Macro family} & \textbf{Slice} & \textbf{Name} & \textbf{Short interpretation} \\
\midrule

\textbf{Abstention}
& A1 & Confident unsupported answer
& Model answers confidently despite no supporting evidence. \\

& A2 & Partial-evidence overconfidence
& Model answers fully when evidence only partially supports the claim. \\

& A3 & Ambiguous forced answer
& Model picks one interpretation instead of flagging ambiguity. \\

\midrule

\textbf{Retrieval}
& R1 & Partial coverage
& Retrieved chunks cover only part of what the question requires. \\

& R2 & Noisy retrieval
& Retrieved chunks contain irrelevant or distracting content. \\

& R3 & Evidence ignored
& Relevant chunks are retrieved but not used in the answer. \\

& R4 & Fragmented evidence
& Relevant information is split across chunks, none sufficient alone. \\

& R5 & Multi-part under-retrieval
& Retriever returns chunks for only one part of a multi-faceted question. \\

& R6 & Distractor-dominated
& High-scoring irrelevant chunks crowd out relevant ones. \\

& R7 & Sparse evidence
& Corpus lacks sufficient content to answer the question. \\

\midrule

\textbf{Grounding}
& G1 & Temporal misattribution
& Correct fact assigned to the wrong time period. \\

& G2 & Entity substitution
& Correct relationship stated with the wrong entity name. \\

& G3 & Unsupported causal bridge
& Model infers a causal link not stated in the context. \\

& G4 & Omitted qualifier
& Context is hedged but the answer states the claim as absolute. \\

& G5 & Broken multi-hop
& Error introduced while chaining reasoning steps across chunks. \\

& G6 & Unsupported synthesis
& Model combines facts across chunks in an unsupported way. \\

& G7 & Citation drift
& Answer uses correct content but attributes it to the wrong entity. \\

\midrule

\textbf{Generation}
& S1 & Partial aspect coverage
& Answer addresses some but not all aspects of the question. \\

& S2 & Shallow summarization
& Answer is too surface-level and misses important context details. \\

& S3 & Underused evidence
& Relevant context is retrieved but not incorporated in the answer. \\

& Q1 & Rambling
& Answer is verbose, unfocused, or contains unnecessary content. \\

& Q2 & Poor structure
& Answer is hard to follow due to disorganized presentation. \\

& Q3 & Internal inconsistency
& Answer contradicts itself within the same response. \\

\bottomrule
\end{tabularx}
\caption{\textsc{Rectify} failure taxonomy: four macro families and 23 fine-grained repair slices.}
\label{tab:failure-taxonomy}
\end{table*}

Cases not removed by these pre-filters are assigned to a primary macro failure family in the following priority order: \emph{abstention}, \emph{retrieval}, \emph{grounding}, and \emph{generation}. This ordering is conservative: unsupported confident answers are handled first, and retrieval is checked before grounding because faithfulness is difficult to interpret when evidence is weak or noisy. Generation is used when retrieval and grounding are adequate, but the final answer remains unsatisfactory. When multiple failure conditions hold, \textsc{Rectify} also records a secondary family to support compound repairs. Within the selected macro family, \textsc{Rectify} assigns the case to a fine-grained repair slice. The current taxonomy defines 23 slices across the four families, with each slice linked to a repair-card template. This repair step is more specific than family-level diagnosis alone: for example, different retrieval failures may require reranking, broader evidence retrieval, or chunking changes, while grounding failures may require more evidence-constrained prompting. The full failure taxonomy is presented in Table~\ref{tab:failure-taxonomy}.

\subsection{Repair-Card Generation}

For each populated repair slice, \textsc{Rectify} creates a repair card: a structured, editable proposal that describes the diagnosed failure, the affected cases, and the suggested configuration-level intervention. Card generation is deterministic. Given the assigned repair slice and affected case IDs, \textsc{Rectify} selects the corresponding template and fills in the target pipeline stage, proposed change, expected benefit, expected tradeoff, and repair scope. Repair cards turn recurring failure patterns into concrete repair hypotheses. For example, a noisy-retrieval slice may suggest enabling reranking, while an underused-evidence slice may suggest a more evidence-constrained prompt. For compound failures, \textsc{Rectify} can combine interventions across stages, such as reranking together with grounded prompting.



\subsection{Human Approval \& Provenance}

\textsc{Rectify} keeps the human at the decision point. The user can inspect the repair slice, review affected cases, edit the proposed parameters, choose the repair scope, and approve or reject the card. Each approval or rejection is stored in a provenance log containing the repair-card identifier, slice type, affected cases, approved parameters, scope, user notes, timestamp, expected benefit, and expected tradeoff. This log makes the repair process auditable: it records what was proposed, what was approved or rejected, which cases were affected, and under which configuration the repair was considered or tested.

\subsection{Optional Sandbox Verification}

After approval, a repair card can be verified in a sandbox before it is accepted as useful. The sandbox applies the approved configuration patch to the affected cases, reruns the RAG pipeline, reevaluates the new outputs, and compares them with the original evaluation results. In our experiments, this verification uses RAGVue as the primary evaluator. When sandbox verification is run, the delta report summarizes whether each affected case improved, remained unchanged, or regressed. \textsc{Rectify} reports aggregate counts, improvement rate, average metric deltas, and per-case before and after outputs. This makes repair proposals empirically checkable rather than merely plausible.

\paragraph{Local Streamlit Application.}
 For interactive use, \textsc{Rectify} provides a Python-based Streamlit interface that exposes the main workflow through a local browser application. Users can clone the repository and start the interface with a standard command such as \texttt{streamlit run streamlit\_app.py}. The local UI supports loading evaluation files, inspecting diagnosed cases, reviewing repair cards, and generating reports without writing code. This interface is intended for practitioners who prefer a point-and-click workflow while keeping data and credentials on their own machine. Selected screenshots of the interface are provided in Appendix~\ref{app:ui}, and the full walkthrough is included in the repository and demo video.
 
\section{Evaluation}

\subsection{Experimental Setup}

We evaluate \textsc{Rectify} on a controlled 100-question synthetic RAG benchmark; dataset construction and question statistics are described in Appendix~\ref{app:dataset}. We test three retriever configurations over the same corpus and question set: BM25 keyword retrieval \citep{10.1561/1500000019}, dense retrieval with \texttt{all-MiniLM-L6-v2} embeddings \citep{Reimers2019SentenceBERTSE,wang2020minilm}, and a hybrid BM25+dense retriever. For all configurations, Mistral-7B \citep{Jiang2023Mistral7} is used as the generator model, served locally via Ollama\footnote{\url{https://ollama.com/}}. The generated outputs are then evaluated with 12 RAGVue \citep{murugaraj2026ragvue} metrics using the same Mistral-7B model as the judge.

\subsection{Results}
We focus the evaluation on three main questions, described below. Additional ablations on the pre-filtering stage and taxonomy depth are reported in Appendix~\ref{app:ablations}.

\begin{table}[t]
\centering
\small
\setlength{\tabcolsep}{5pt}
\renewcommand{\arraystretch}{1.10}
\begin{tabular}{@{}lccc@{}}
\toprule
\textbf{Measure} & \textbf{BM25} & \textbf{Dense} & \textbf{Hybrid} \\
\midrule
\multicolumn{4}{c}{\textbf{\textit{Case accounting}}} \\
\midrule
Unanswerable prefiltered & 16 & 16 & 16 \\
Gold-answer filter & 62 & 75 & 71 \\
\cmidrule(lr){1-4}
\textit{Subtotal: filtered before diagnosis} & \textit{78} & \textit{91} & \textit{87} \\
\cmidrule(lr){1-4}
No repair family triggered & 6 & 4 & 7 \\
Final repair agenda & 16 & \textbf{5} & 6 \\
\addlinespace[2pt]
\midrule
\multicolumn{4}{c}{\textbf{\textit{Repair-slice breakdown}}} \\
\midrule
R2: Noisy retrieval & 7 & 0 & 0 \\
R5: Multi-part under-retrieval & 3 & 4 & 1 \\
S3: Underused evidence & 6 & 1 & 5 \\
\bottomrule
\end{tabular}
\caption{\textsc{Rectify} case accounting and repair-slice breakdown across BM25, Dense, and Hybrid retrieval. The upper block shows filtering and routing outcomes; the lower block shows the final repair agenda.}
\label{tab:rectify-rq1}
\end{table}

\paragraph{How do failure profiles differ across retriever configurations, and does \textsc{Rectify} surface consistent and interpretable diagnostics?}
Table~\ref{tab:rectify-rq1} summarizes the main \textsc{Rectify} outputs for the three retriever configurations: how cases are filtered before diagnosis, how many cases remain in the final repair agenda, and which repair slices are triggered. The results show that retriever choice substantially changes the repair agenda. BM25 leaves 16 cases requiring repair, while dense and hybrid retrieval reduce this to 5 and 6 cases, respectively. This indicates that many failures are retrieval-driven: semantic retrieval resolves cases where BM25 retrieves lexically plausible but weakly relevant evidence. Hybrid retrieval does not clearly dominate dense retrieval in the final repair agenda, but it still preserves complementary lexical and semantic signals. The repair-slice breakdown shows why aggregate evaluator scores (Table~\ref{tab:ragvue-metrics-app}) alone are not sufficient. Under BM25, the main repair need is R2 noisy retrieval. With dense and hybrid retrieval, R2 drops to 0 cases, and the remaining failures shift toward R5 multi-part under-retrieval and S3 underused evidence. These failure types can all lead to weak faithfulness, completeness, or answer relevance, but they require different repairs: reranking for R2, broader evidence retrieval for R5, and prompt-level or abstention-policy changes for S3. This demonstrates that \textsc{Rectify} surfaces interpretable diagnostics that are directly tied to repair actions.

\begin{table}[t]
\centering
\small
\setlength{\tabcolsep}{4pt}
\renewcommand{\arraystretch}{1.10}
\begin{tabular}{@{}p{2.8cm}cp{3.5cm}@{}}
\toprule
\textbf{Pattern across retrievers} & \textbf{Count} & \textbf{Interpretation} \\
\midrule
No repair in all retrievers & 81 & The question is handled correctly across BM25, Dense, and Hybrid retrieval \\
Fails in at least one retriever & 19 & Questions used to analyze repair-slice consistency \\
\midrule
Same S3 slice in $\geq$2 retrievers & 3 & Evidence is retrieved but underused by the generator \\
Same R5 slice in $\geq$2 retrievers & 2 & The question needs evidence from multiple documents \\
R2 only with BM25 & 7 & BM25 introduces noisy keyword-matched chunks \\
Other one-off or mixed failures & 7 & Failure depends on retriever setting or slice interaction \\
\bottomrule
\end{tabular}
\caption{Repair-slice consistency across BM25, Dense, and Hybrid retrieval.}
\label{tab:slice-consistency}
\end{table}

\paragraph{Are repair slices consistent across retrievers?}
We compare how \textsc{Rectify} labels the same 100 questions under BM25, Dense, and Hybrid retrieval. If the same question receives the same repair slice under multiple retrievers, we treat it as a stable failure pattern. If a slice appears only under one retriever, the failure is more likely tied to that retrieval configuration. Table~\ref{tab:slice-consistency} shows that 81 questions need no repair under any retriever. Among the 19 questions that fail at least once, repeated S3 and R5 labels reveal stable problems: S3 means that retrieved evidence is available but underused, while R5 means that the question requires evidence from multiple documents. In contrast, seven R2 cases appear only with BM25, indicating keyword-matching noise that disappears with dense or hybrid retrieval. This suggests that \textsc{Rectify}'s slices are not arbitrary labels: they help separate stable failure patterns from retriever-specific errors.

\begin{figure}[t]
\centering
\includegraphics[width=\linewidth]{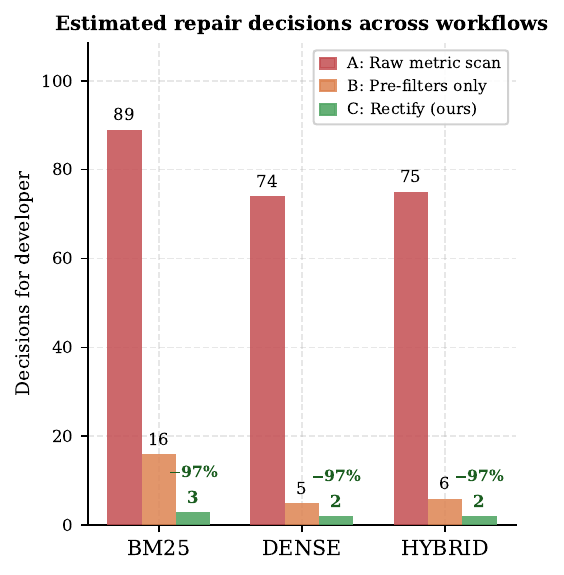}
\caption{Estimated debugging effort across three workflows: raw metric scan, pre-filtering only, and \textsc{Rectify} with pre-filtering plus repair-slice grouping. Effort is measured as the number of case-level or card-level repair decisions.}
\label{fig:debugging-effort}
\end{figure}

\paragraph{Does \textsc{Rectify} reduce debugging effort?}
We estimate debugging effort under three workflows. In a raw metric-scan workflow, a developer inspects every case where a core metric falls below a threshold. With pre-filtering only, the developer inspects the post-filter-failing cases individually, without any grouping. With \textsc{Rectify}, the same pre-filters are applied first, and the remaining failures are then grouped into repair cards, so the developer reviews one card per repair slice and makes one repair decision per cluster. Figure~\ref{fig:debugging-effort} shows that the raw metric scan requires 74-89 individual case inspections per retriever configuration. The pre-filters reduce this to 5-16 post-filter cases. \textsc{Rectify}'s taxonomy then groups those cases into only 2-3 repair cards while still covering the full final repair agenda. Under our decision-count estimate, \textsc{Rectify} reduces the number of repair decisions by 97\%. This reduction is important because it changes the debugging unit. Instead of making many ad hoc case-level judgements, the developer reviews a small number of cluster-level repair hypotheses: for example, enabling reranking for R2 noisy retrieval, increasing top-$k$ for R5 multi-part under-retrieval, or adjusting prompting for S3 underused evidence. Finally, because sandbox verification is optional in the workflow, we report detailed sandbox results separately in Appendix~\ref{app:sandbox-verification}. These results illustrate how approved repair cards can be checked through before and after deltas before a developer accepts or rejects a repair.

\section{Conclusion}
We presented \textsc{Rectify}, an interactive workbench prototype for moving from RAG evaluation to diagnosis, repair, and verification.  Our experiments show that \textsc{Rectify} surfaces actionable failure profiles across retriever configurations. As retrieval quality improves, the repair agenda becomes smaller, and the dominant failure modes shift: BM25 mainly suffers from noisy retrieval, while dense and hybrid retrieval expose more specific failures, such as multi-part under-retrieval and underused evidence. This shift would be difficult to interpret from aggregate metrics alone, but becomes visible through \textsc{Rectify}'s slice-level diagnosis. The results also show why repair should be treated as a review-and-verification workflow rather than an automatic patching step. Pre-filtering reduces unnecessary repair workload, fine-grained slicing produces targeted repair cards, and sandbox verification exposes both successful repairs and cases where a proposed intervention should be rejected or revised. Overall, \textsc{Rectify} contributes a post-evaluation workflow for RAG development: diagnose recurring failures, propose inspectable repairs, keep the developer in control, and verify changes with before and after evidence.

\section*{Limitations and Future Work}

\textsc{Rectify} is a local, interactive workbench for post-evaluation RAG repair, not an automatic production optimizer. Its current implementation uses a deterministic taxonomy with 23 repair slices. This design makes routing decisions reproducible and auditable, while also making the taxonomy easy to extend as new domains, retrievers, or generation failure patterns introduce additional repair needs. The pre-filtering stage can use benchmark annotations such as gold answers and answerability labels when they are available, but these annotations are not required for the core workflow: without them, \textsc{Rectify} still performs diagnosis from evaluator signals and simply skips annotation-based filtering. Sandbox verification reruns affected cases and reevaluates the new outputs, so developers can apply it selectively to high-impact repair cards or use it as a final check before accepting a proposed configuration change.

\textcolor{black}{Future work will expand \textsc{Rectify} along three directions: evaluating it on larger real-world knowledge bases, extending the repair taxonomy to additional RAG architectures and application domains, and studying how the workflow transfers across evaluators with different diagnostic signals. We also plan to test additional generator-judge combinations to further assess the stability of the observed repair profiles.}

\section*{Ethics Statement}

\textsc{Rectify} is a developer-facing workbench for post-evaluation RAG diagnosis and repair. It does not automatically modify or deploy RAG systems; repair cards require human review, approval, and optional sandbox verification. Our experiments use a synthetic benchmark with fictional entities and do not involve private or personally identifiable data. In real deployments, users should ensure that input documents, evaluation files, model outputs, and connected services comply with relevant privacy, licensing, and data-governance requirements. \textsc{Rectify}'s suggestions should be treated as decision support rather than guaranteed fixes, especially in high-stakes domains.

\bibliography{custom}

\clearpage

\appendix

\section{Additional Screenshots}
\label{app:ui}

\begin{figure*}[t]
\centering
\begin{tabular}{ccc}
\includegraphics[width=0.31\textwidth]{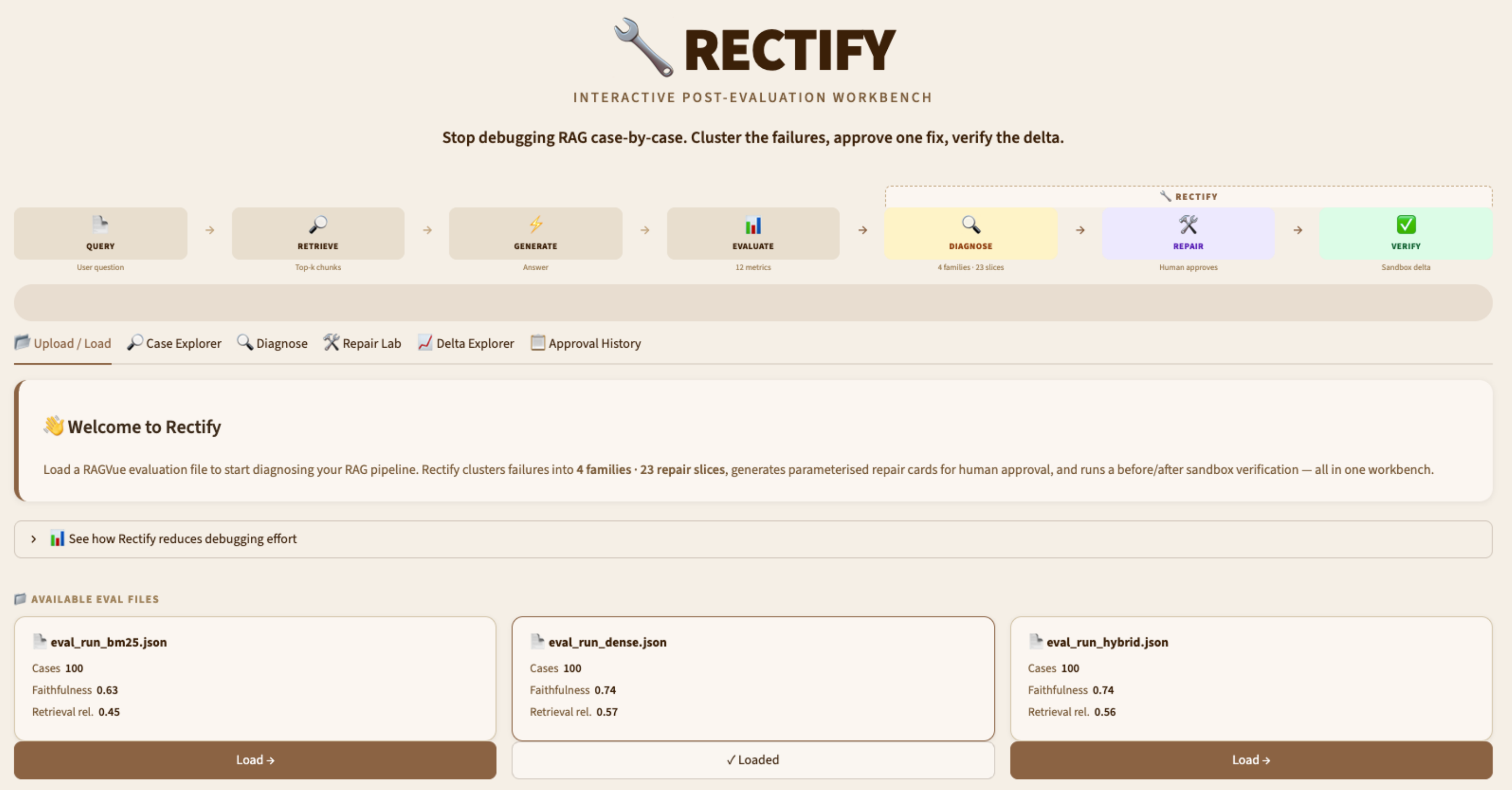} &
\includegraphics[width=0.31\textwidth]{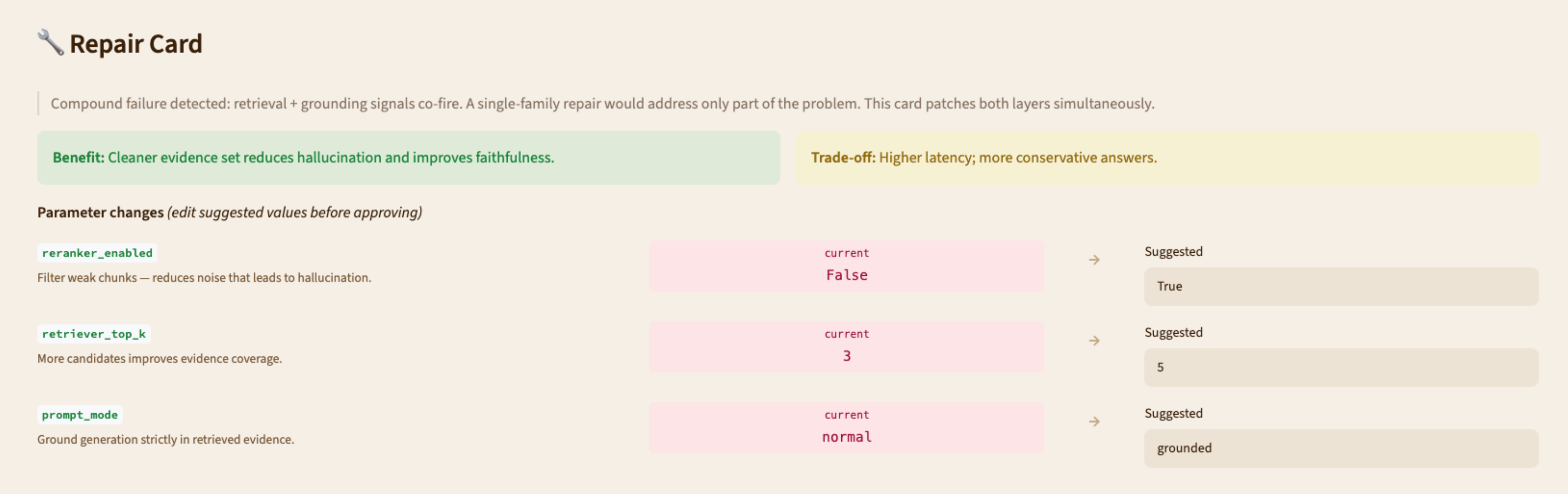} &
\includegraphics[width=0.31\textwidth]{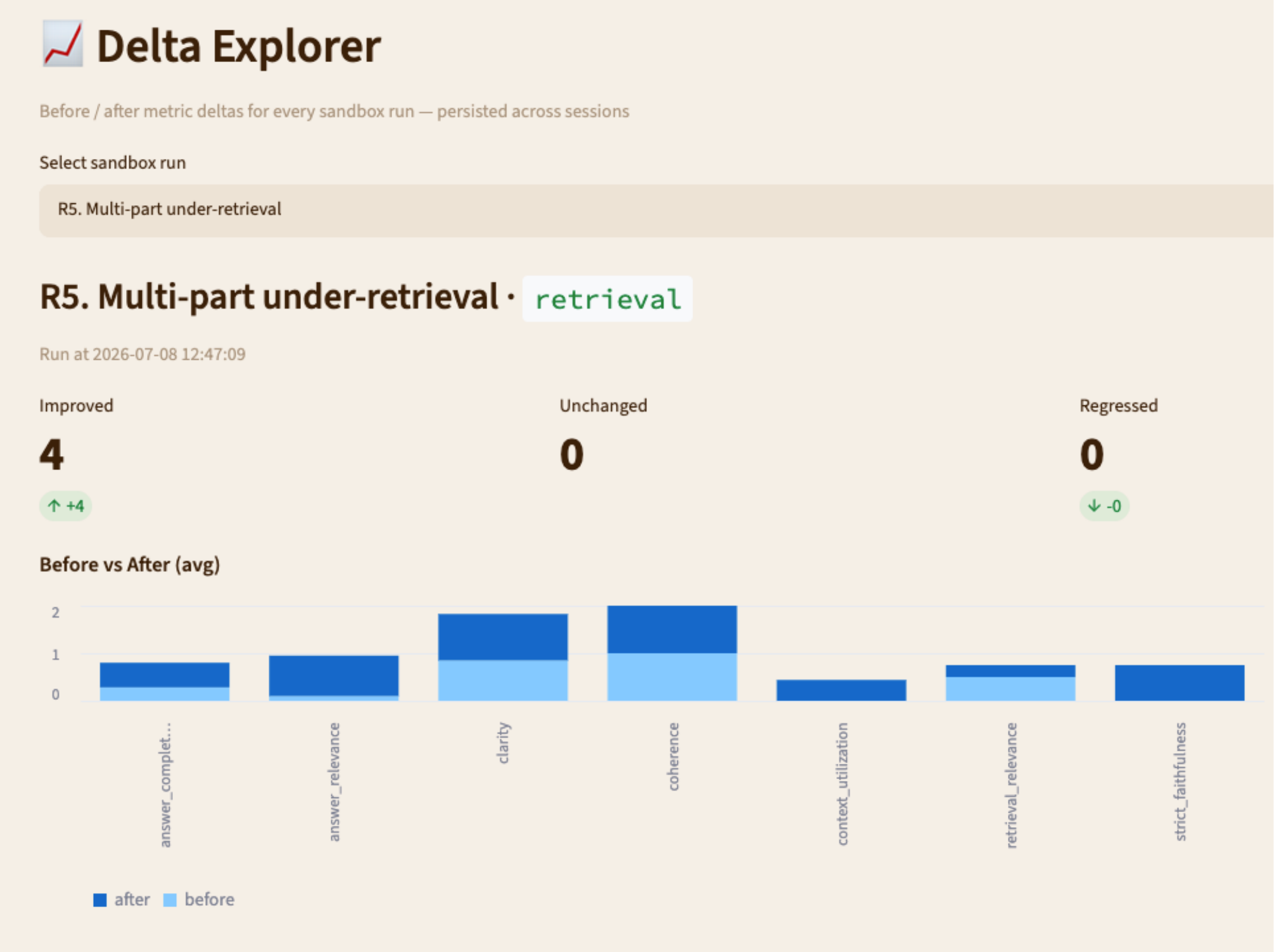} \\
(a) Home & (b) Repair card (c) Delta explorer
\end{tabular}
\caption{Selected screenshots of the local Python-based \textsc{Rectify} Streamlit interface. }
\label{fig:ui-screenshots}
\end{figure*}

Example UI screenshots are shown in Figure~\ref{fig:ui-screenshots}, and the full walkthrough is available in the repository and the demo video.

\section{Dataset}
\label{app:dataset}

We construct a controlled synthetic corpus with 30 short documents about 10 fictional companies and associated entities. The documents cover company profiles, persons, products, events, and thematic summaries and encode factual relations such as founding year, headquarters, acquisitions, partnerships, and flagship products. We generate 100 questions across eight types: factoid (25), multi-part (20), multi-hop (16), temporal (13), comparison (10), explicitly unanswerable (10), multi-hop unanswerable (4), and temporal unanswerable (2). In total, 84 questions are answerable, and 16 are unanswerable. Answerable questions include gold answers and known relevant document identifiers; unanswerable questions test whether the system correctly abstains when the required evidence is absent from the corpus.

\paragraph{Why synthetic?}
We use a small controlled corpus because \textsc{Rectify} is evaluated as a repair workbench, not as an open-domain QA system. Repair evaluation requires more control than gold answers alone: we need known relevant documents, controlled unanswerable cases, and known corpus coverage gaps to distinguish retrieval failures, missing evidence, poor evidence use, and correct abstention. The synthetic setup also lets us create targeted failure conditions, such as noisy retrieval, missing multi-hop evidence, underused evidence, and unsupported answers. We acknowledge that this corpus is narrower than real enterprise settings, and future work will evaluate \textsc{Rectify} on larger real-world knowledge bases.

\section{Full RAGVue Metric Scores}
\label{app:ragvue-metrics}

Table~\ref{tab:ragvue-metrics-app} reports the full set of RAGVue metric scores used as evaluator signals in the cross-retriever analysis. 
\begin{table}[t]
\centering
\scriptsize
\setlength{\tabcolsep}{4.5pt}
\renewcommand{\arraystretch}{1.10}
\begin{tabular}{@{}lccc@{}}
\toprule
\textbf{RAGVue metric} & \textbf{BM25} & \textbf{Dense} & \textbf{Hybrid} \\
\midrule
Strict faithfulness & 0.627 & \textbf{0.745} & 0.741 \\
Retrieval relevance & 0.450 & \textbf{0.567} & 0.563 \\
Retrieval coverage & 0.757 & 0.823 & \textbf{0.830} \\
Answer completeness & 0.391 & \textbf{0.474} & 0.465 \\
Answer relevance & 0.648 & \textbf{0.775} & 0.762 \\
Context utilization & 0.465 & 0.570 & \textbf{0.596} \\
Multi-hop faithfulness & 0.655 & \textbf{0.778} & 0.762 \\
Coherence & 0.999 & \textbf{1.000} & \textbf{1.000} \\
Clarity & 0.971 & 0.981 & \textbf{0.987} \\
Negative rejection & \textbf{0.990} & 0.970 & \textbf{0.990} \\
Answer conciseness & 0.975 & \textbf{0.993} & 0.991 \\
Implicit contradiction & \textbf{0.924} & 0.909 & 0.913 \\
\bottomrule
\end{tabular}
\caption{Full RAGVue metric scores for BM25, Dense, and Hybrid retrieval over the same 100-question benchmark. Bold marks the best value in each row.}
\label{tab:ragvue-metrics-app}
\end{table}




\section{Sandbox Verification Results}
\label{app:sandbox-verification}

Table~\ref{tab:sandbox-summary} illustrates sandbox verification for two representative repair slices. For R2 noisy retrieval, applying reranking to seven BM25 cases improves all seven. For R5 multi-part under-retrieval, increasing top-$k$ from 3 to 6 improves five of eight cases, leaves two unchanged, and regresses one. These results support the review-and-verification loop in \textsc{Rectify}. Repair cards are testable hypotheses, not guaranteed fixes: sandbox deltas help developers decide whether a proposed repair improves, leaves unchanged, or regresses the affected cases before accepting it.

\begin{table}[t]
\centering
\scriptsize
\setlength{\tabcolsep}{4pt}
\renewcommand{\arraystretch}{1.10}
\begin{tabular}{@{}p{0.8cm}p{2.1cm}cccc@{}}
\toprule
\textbf{Slice} & \textbf{Repair} & \textbf{n} & \textbf{Improved} & \textbf{Unchanged} & \textbf{Regressed} \\
\midrule
R2 & Enable reranker & 7 & \textbf{7} & 0 & 0 \\
R5 & Increase top-$k$ 3$\rightarrow$6 & 8 & 5 & 2 & 1 \\
\bottomrule
\end{tabular}
\caption{Sandbox verification results for two representative repair slices.}
\label{tab:sandbox-summary}
\end{table}

\section{Additional Ablations}
\label{app:ablations}

\subsection{Pre-filters Reduce Unnecessary Repair Candidates}

\textsc{Rectify} applies two pre-filters before failure routing, as described in Section~\ref{main:failure_family}. These filters are not intended as a replacement for evaluation; they reduce unnecessary repair work before the remaining cases are diagnosed and grouped into repair slices. Table~\ref{tab:prefilter-ablation} shows the effect of these filters. A raw metric scan flags 89, 74, and 75 cases for BM25, dense, and hybrid retrieval, respectively. When the failure taxonomy is applied without pre-filters, the candidate set becomes 68, 54, and 55 cases. With the abstention and gold-answer filters enabled, the final repair agenda drops to 16, 5, and 6 cases. This corresponds to an 82--93\% reduction compared with raw metric scanning and a 77--91\% reduction compared with taxonomy-based routing without pre-filters. The number of active failure slices also decreases from eight to three, two, and two, producing a smaller and more focused repair agenda.

\begin{table}[t]
\centering
\scriptsize
\setlength{\tabcolsep}{4pt}
\renewcommand{\arraystretch}{1.10}
\begin{tabular}{@{}lccc@{}}
\toprule
\textbf{Measure} & \textbf{BM25} & \textbf{Dense} & \textbf{Hybrid} \\
\midrule
Raw metric scan & 89 & 74 & 75 \\
Family routing, no pre-filters & 68 & 54 & 55 \\
\textsc{Rectify} with pre-filters & \textbf{16} & \textbf{5} & \textbf{6} \\
\cmidrule(lr){1-4}
Reduction vs.\ raw scan & 82\% & 93\% & 92\% \\
Reduction vs.\ no pre-filters & \textbf{77\%} & \textbf{91\%} & \textbf{89\%} \\
\midrule
\end{tabular}
\caption{Ablation of \textsc{Rectify}'s pre-filtering stage. Raw metric scan counts broadly flagged cases, while family routing applies the taxonomy before disabling the abstention and gold-answer filters.}
\label{tab:prefilter-ablation}
\end{table}

\begin{table}[t]
\centering
\scriptsize
\setlength{\tabcolsep}{4pt}
\renewcommand{\arraystretch}{1.10}
\begin{tabular}{@{}lccc@{}}
\toprule
\textbf{Measure} & \textbf{BM25} & \textbf{Dense} & \textbf{Hybrid} \\
\midrule
\multicolumn{4}{c}{\textit{Family-only routing}} \\
\midrule
Repair cards & 2 & 2 & 2 \\
Avg.\ cases per card & 8 & 3 & 3 \\
Cases merged into retrieval card & 10 & 4 & 1 \\
Cases merged into generation card & 6 & 1 & 5 \\
\midrule
\multicolumn{4}{c}{\textit{Fine-grained slices}} \\
\midrule
Repair cards & 3 & 2 & 2 \\
Avg.\ cases per card & 5 & 3 & 3 \\
R2/R5 separated & Yes & -- & -- \\
S3 isolated from retrieval & Yes & Yes & Yes \\
\bottomrule
\end{tabular}
\caption{Ablation of taxonomy depth over the post-filter repair agenda. Fine-grained slices separate failures that require different repair actions.}
\label{tab:taxonomy-depth-ablation}
\end{table}

\subsection{Fine-Grained Slices Produce More Targeted Repair Cards}

We ablate the second layer of \textsc{Rectify}'s taxonomy by comparing fine-grained repair slices with a family-only routing baseline. Family-only routing groups failures into broad families such as retrieval or generation, while fine-grained routing separates them into repair slices such as R2 noisy retrieval, R5 multi-part under-retrieval, and S3 underused evidence. Table~\ref{tab:taxonomy-depth-ablation} shows the practical effect of this distinction. Under BM25, family-only routing merges 10 retrieval-family cases into a single card, although these cases require different interventions: R2 points to reranking, while R5 points to broader evidence retrieval, such as increasing top-$k$. Fine-grained slicing separates these cases into more specific repair cards and keeps S3 underused-evidence cases separate from retrieval failures. This makes the repair agenda more actionable: developers review slightly more specific cards, but each card corresponds to a clearer repair hypothesis.


\end{document}